\documentclass[final]{svjour3}
\usepackage{graphicx}
\usepackage{subcaption}
\usepackage{rotating}
\usepackage{amssymb}
\usepackage{mathptmx}
\usepackage[numbers]{natbib}
\makeatletter
\journalname{Journal of Low Temperature Physics}
\bibpunct{}{}{,}{s}{}{,}

\begin{document}

\setcitestyle{square}
\setcitestyle{numbers}

\newcommand{\hdblarrow}{H\makebox[0.9ex][l]{$\downdownarrows$}-}
\title{Stacked Wafer Gradient Index Silicon Optics with Integral Anti-reflection Layers}

\author{F. Defrance, G. Chattopadhyay, J. Connors, S. Golwala, M. I. Hollister, C. Jung-Kubiak , E. Padilla, S. Radford, J. Sayers , E. C. Tong, H. Yoshida}

\institute{Division of Physics, Mathematics and Astronomy, California Institute of Technology,\\ Pasadena, CA 91125, US\\ 
\email{fdefranc@caltech.edu}}

\authorrunning{F. Defrance et al.}
\maketitle

\begin{abstract}

Silicon optics with wide bandwidth anti-reflection (AR) coatings, made of multi-layer textured silicon surfaces, are developed for millimeter and submillimeter wavelengths. Single and double layer AR coatings were designed for an optimal transmission centered on 250~GHz, and fabricated using the DRIE (Deep Reaction Ion Etching) technique. Tests of high resistivity silicon wafers with single-layer coatings between 75~GHz and 330~GHz are presented and compared with the simulations.


\end{abstract}

\section{Introduction}

Observations at millimeter and submillimeter wavelengths (tens of GHz to THz frequencies) are central to addressing a range of forefront topics in astronomy. Studies of the polarization of the Cosmic Microwave Background radiation (CMB) and of the structure, internal motions, and evolution of galaxy clusters, share the primary technical requirements of high sensitivity and wide field imaging across a wide spectral bandwidth. Achromatic transmission optics, including lenses and windows, are a critical technology. Although the requirements of future astronomy missions (on the ground and in space) are the primary motivation for developing this technology, other applications include studies of planets and other solar system objects, monitoring the earth's atmosphere, and emerging reconnaissance and security uses.
Silicon's high refractive index and low loss make it an ideal optical material for developing achromatic transmission optics at these frequencies. It is even possible to use silicon for ambient temperature vacuum windows. Silicon's large refractive index, however, necessitates anti-reflection (AR) coating. Moreover, multilayer anti-reflection treatments are necessary for wide spectral bandwidths, with wider bandwidths requiring more layers. To this end, we are developing multilayer coatings for silicon by bonding together wafers individually patterned with deep reactive ion etching (DRIE).
While a standard approach to anti-reflection coating is to deposit or laminate dielectric layers of appropriate refractive index \cite{gatesman00}, it is difficult (but not impossible) to find low loss dielectrics with the correct refractive index and other properties to match silicon well, especially if more than one layer is required, operation up to THz frequencies is desired, and/or the optic will be used cryogenically. Textured surfaces are an attractive alternative to dielectric anti-reflection coatings. 
For millimeter wavelengths, multi-layer anti-reflection textures with up to 4:1 bandwidths have been cut successfully into silicon lens surfaces with a dicing saw \cite{datta13,wheeler14} but this technique becomes unusable at frequencies of 300 GHz and higher given the saw dimensions. Laser machining is being explored but demonstrations are not yet available. DRIE works well on flat surfaces (and has been demonstrated for narrowband windows to THz frequencies), but there are limits to the depth and aspect ratio of the features it can create. Furthermore etching has not been adapted to large, curved optics.
We are pursuing a hybrid approach to this problem: construct a silicon optic by stacking flat patterned wafers. The starting point is a multilayer optical design incorporating both an axial gradient in the refractive index for anti-reflection and a radial index gradient for focusing. For each optical layer, a hole or post pattern is used to achieve the required effective index of refraction. Using a novel multilayer etching procedure, several layers of the optical structure are fabricated on a flat wafer. Several individually patterned wafers are stacked and bonded together to produce the completed optic. This approach can thus address the aspect ratio limitations of DRIE, and it obviates etching on curved surfaces.
We present our results to date, which include measurements at 75-330 GHz of single-layer coatings. These measurements indicate the basic technique is sound. Our near-term goal is to produce a 10-cm lens with a 7-layer coating providing 5.5:1 bandwidth from 75 to 420 GHz, eventually scaling up to 15-cm, 30-cm, and larger elements.

\section{Design and Simulation}

\subsection{Effective refractive index}

\begin{figure}[htbp]
  \begin{center}
  \includegraphics[width=8cm]{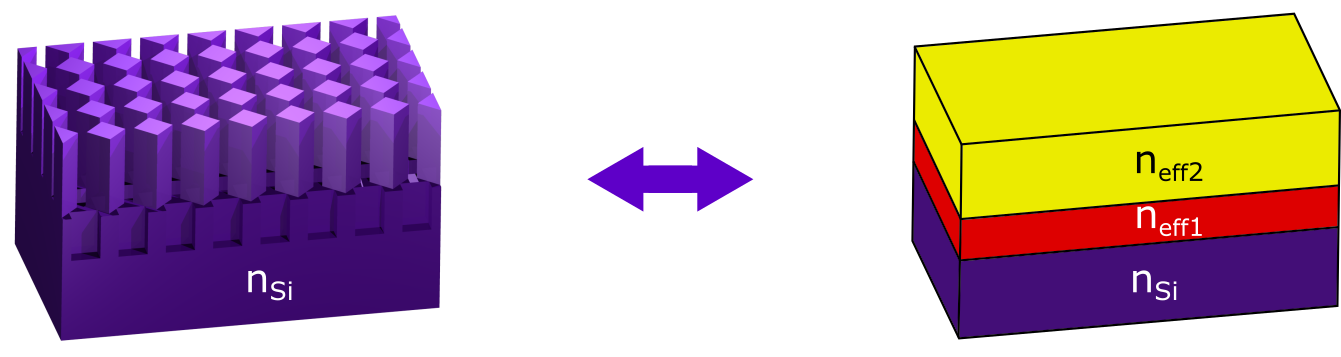}
  \caption{\label{fig:neff3D}Example of a double layer AR structure with two different effective refractive indexes}
  \end{center}
\end{figure}

Texturing an optical surface can reduce reflections by effectively reducing the refractive index of the surface layer. Textured surfaces are an attractive alternative to dielectric AR coatings. Artificial AR structures have been produced for wavelengths from the UV \cite{huang07} to the microwave \cite{lamb96}. So long as the characteristic size of the surface texture is (much) smaller than the wavelength, the AR structure behaves like a continuous medium. By suitable choice of the structure geometry, layers with specific refractive indexes can be fabricated. The refractive index obtained by texturing the surface is called effective refractive index (cf. fig.~\ref{fig:neff3D}).

\begin{figure}[htbp]
  \begin{center}
  \includegraphics[width=8cm]{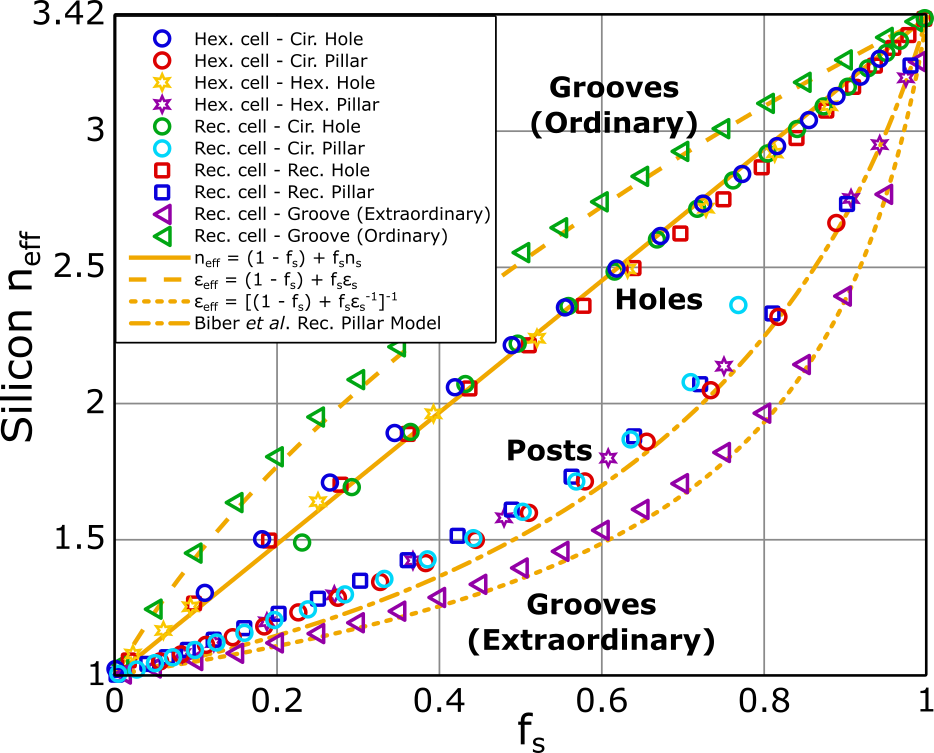}
  \caption{\label{fig:neff}Relation between filling factor ($f_s$) and effective index ($n_{eff}$) for different geometries. HFSS simulations (symbols) and models predictions (lines).}
  \end{center}
\end{figure}

Different pattern geometries give different effective refractive indexes for a same filling factor. Fig~\ref{fig:neff} shows the effective refractive indexes of holes, posts and grooves of different shapes made of (or made in) silicon. Two groove directions were studied, parallel to the electric field (TE or ordinary) and perpendicular to the electric field (TM or extraordinary), as explained by Brundrett et al. \cite{Brundrett94}. We notice that the relation between effective refractive index and filling factor is also very different for posts and holes, but the shape of these posts or holes (square, circular, hexagonal, hexagonal) has little effect. The effective refractive index of posts is detailed by Biber et al. \cite{biber03}.

\subsection{Current design}

We designed and fabricated two single-layer AR coatings (one with square holes and one with square posts), and one double-layer AR coating (with square holes and square posts). These designs were optimized to have a maximum transmission around 250~GHz and the spatial period of the patterns is 125 $\mu m$ (cf. figs.~\ref{fig:1lay_sem} \& \ref{fig:2lay_sem}). 

\begin{itemize}
\item The two single-layer AR coatings were designed with a thickness of 162 $\mu m$ ($\lambda_0/(4.n_{eff})$), where $n_{eff} = 1.85$, and $\lambda_0$ is the wavelength in vacuum (1.2 mm for a frequency of 250~GHz). The side of the holes is 101 $\mu m$ and the side of the posts is 99 $\mu m$.
\item The double-layer AR coating was designed with a top layer of square posts ($n_{eff} = 1.39$, height = 215 $\mu m$ and side = 72 $\mu m$) and a bottom layer of square holes ($n_{eff} = 2.46$, height = 121 $\mu m$ and side = 77 $\mu m$).
\end{itemize}

\section{Fabrication}

\begin{figure*}[th]
    \centering
    \begin{subfigure}[t]{0.4\textwidth}
        \centering
        \includegraphics[height=1.2in]{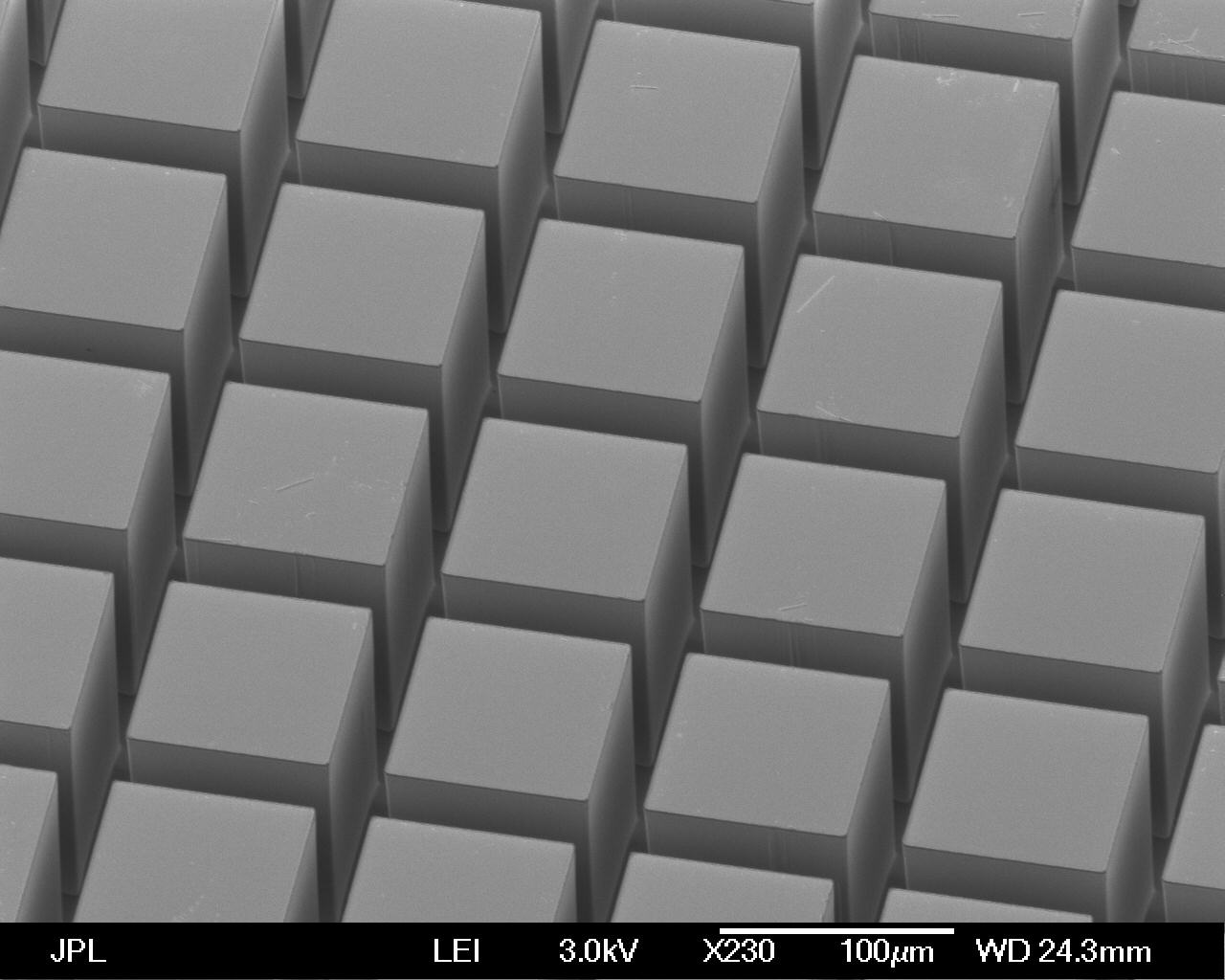}
 %       \caption{Lorem ipsum}
    \end{subfigure}%
    \hspace{0.8cm}
    \begin{subfigure}[t]{0.4\textwidth}
        \centering
        \includegraphics[height=1.2in]{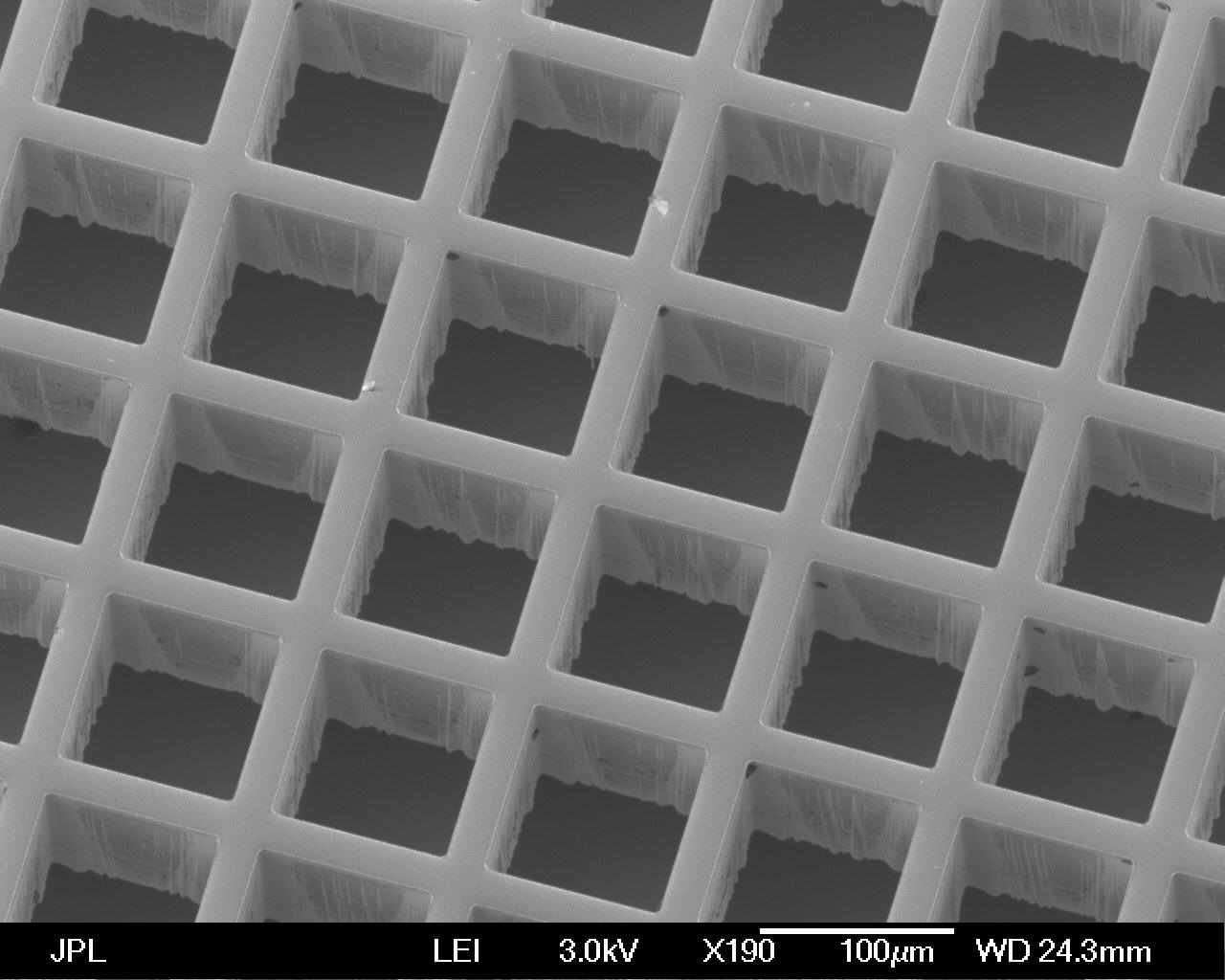}
 %       \caption{Lorem ipsum, lorem ipsum,Lorem ipsum, lorem ipsum,Lorem ipsum}
    \end{subfigure}
    \caption{SEM pictures of fabricated single layer silicon wafers, with $n_{eff} = 1.85$ and a height of 162 $\mu m$. {\it Left,} square posts (side  = 99 $\mu m$) and {\it right,} square holes (side = 101 $\mu m$).}
    \label{fig:1lay_sem}

    \begin{subfigure}[t]{0.28\textwidth}
        \centering
        \includegraphics[height=1.65in]{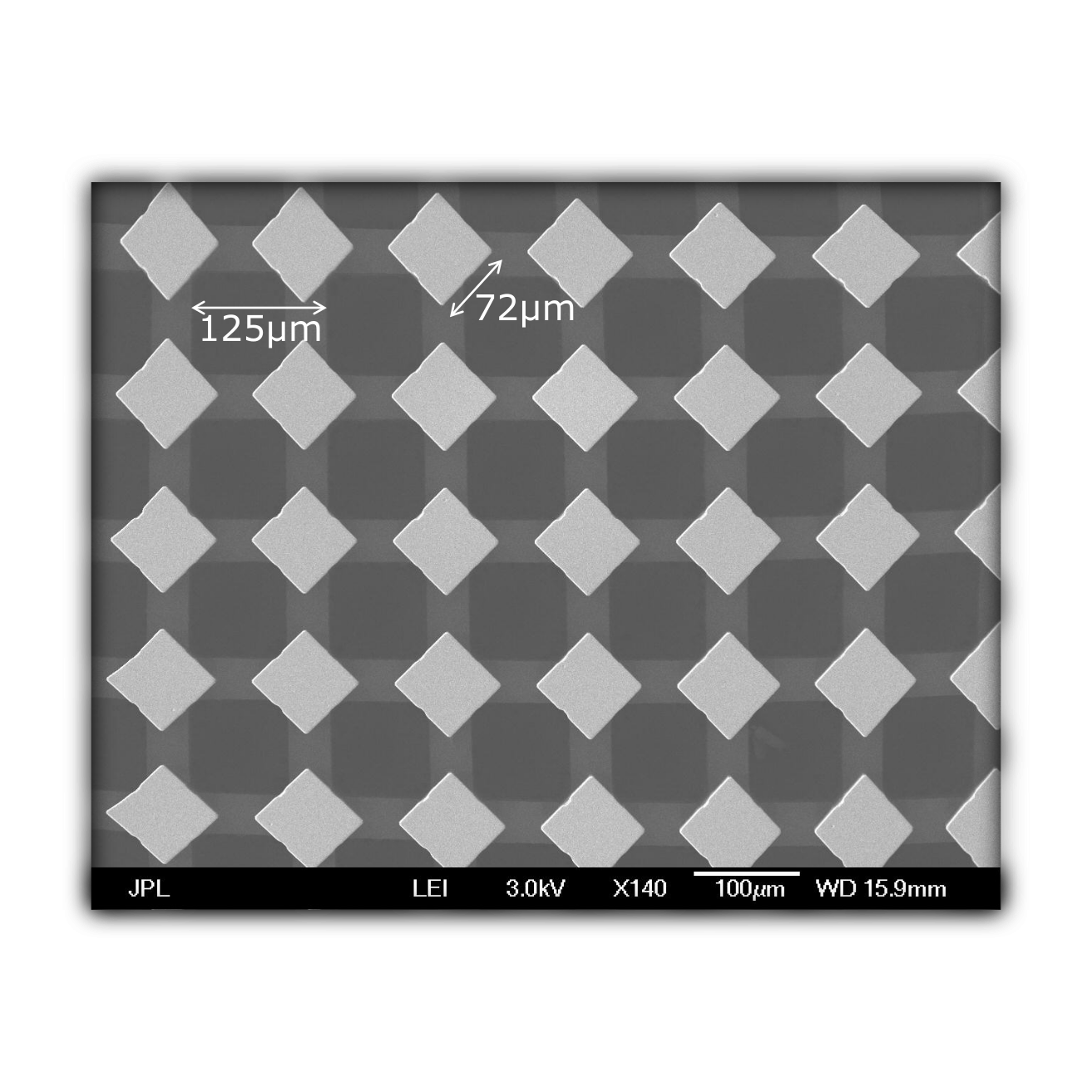}
 %       \caption{Lorem ipsum}
    \end{subfigure}%
    \hspace{0.36cm}
    \begin{subfigure}[t]{0.28\textwidth}
        \centering
        \includegraphics[height=1.65in]{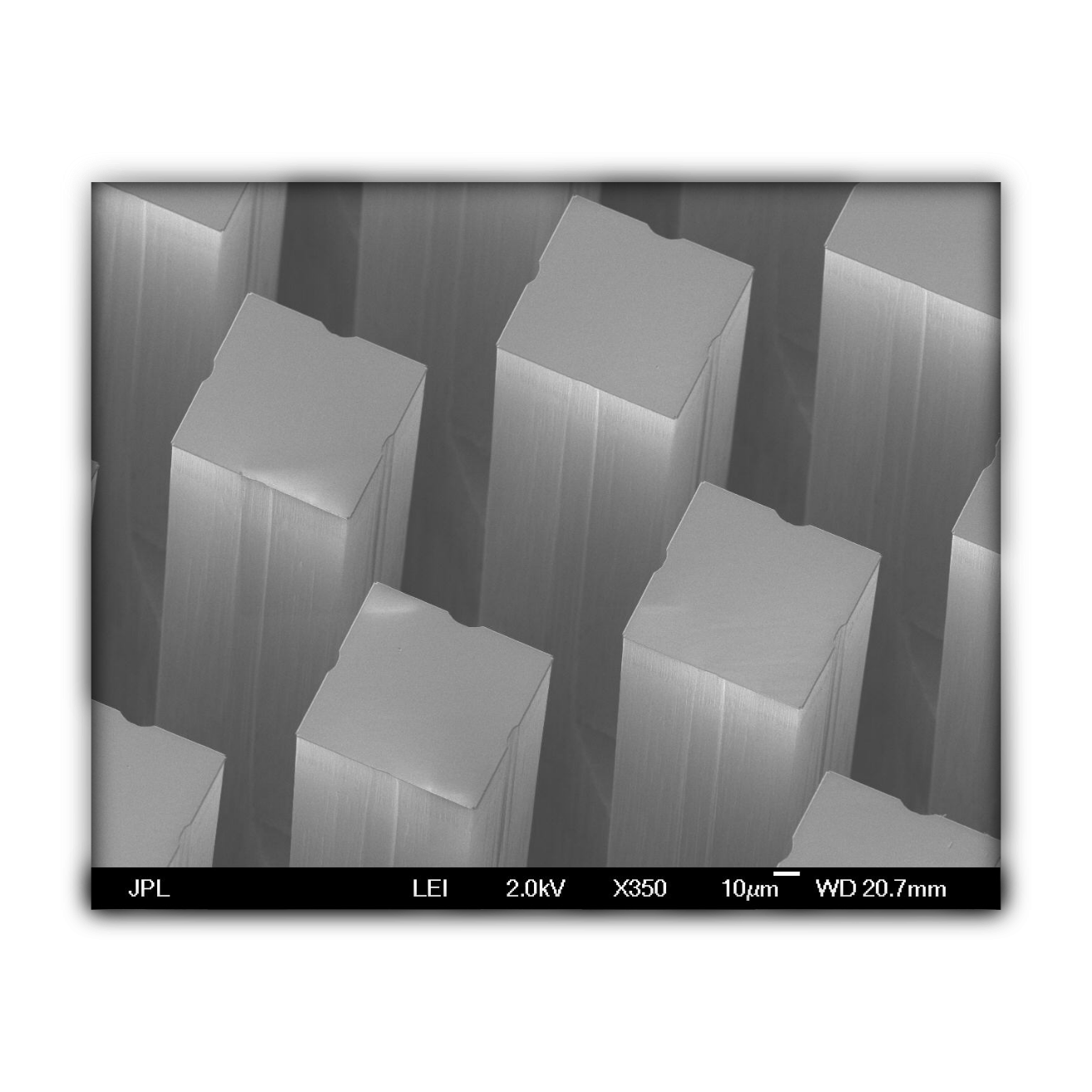}
 %       \caption{Lorem ipsum, lorem ipsum,Lorem ipsum, lorem ipsum,Lorem ipsum}
    \end{subfigure}
    \hspace{0.3cm}
    \begin{subfigure}[t]{0.28\textwidth}
        \centering
        \includegraphics[height=1.65in]{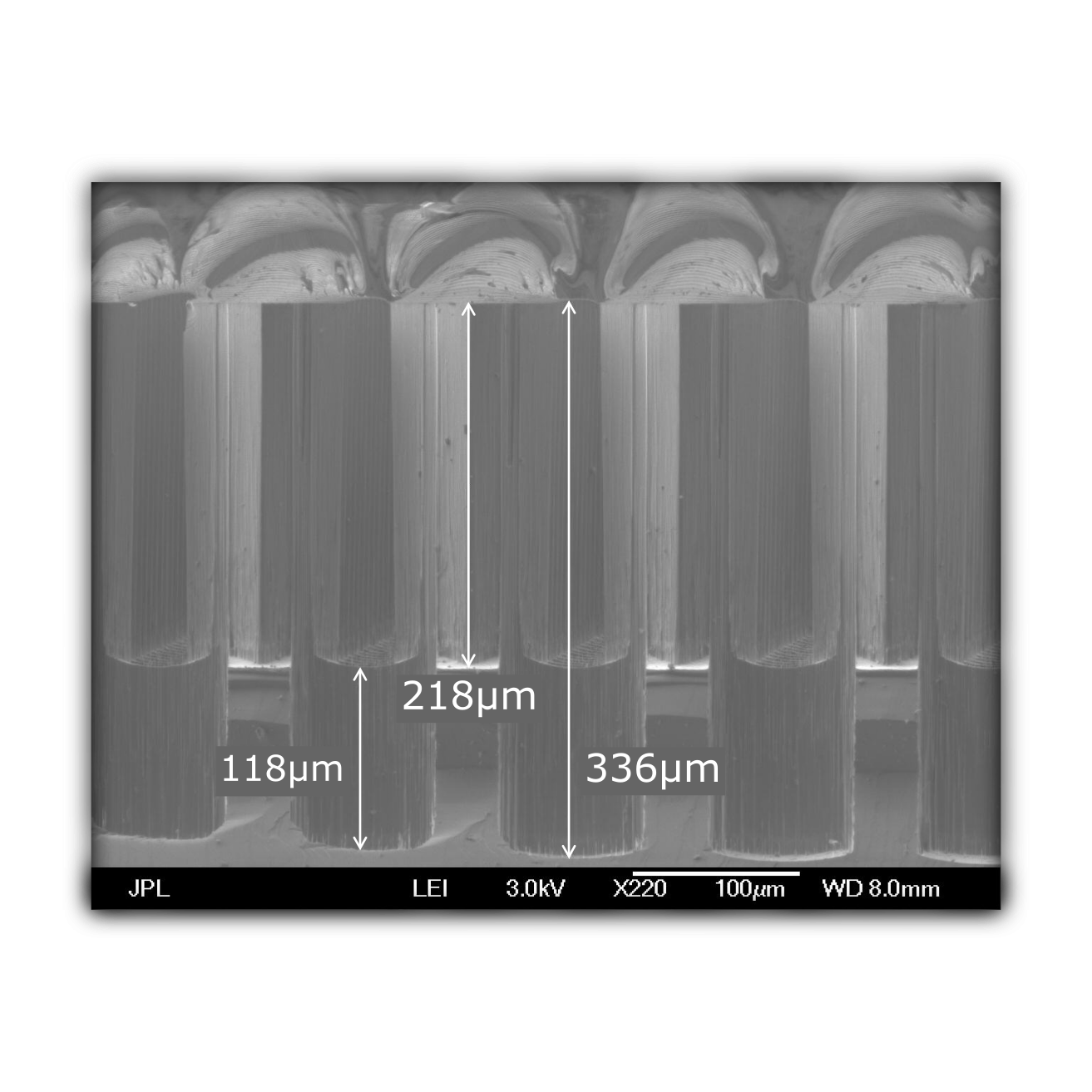}
 %       \caption{Lorem ipsum, lorem ipsum,Lorem ipsum, lorem ipsum,Lorem ipsum}
    \end{subfigure}
    \caption{SEM pictures of fabricated double layer silicon wafers. Top layer: square posts ($n_{eff} = 1.39$, height = 215 $\mu m$ and side = 72 $\mu m$), bottom layer: square holes ($n_{eff} = 2.46$, height = 121 $\mu m$ and side = 77 $\mu m$)}
    \label{fig:2lay_sem}
\end{figure*}

The fabrication of the 1- and 2-layer AR coatings was done at JPL (Jet Propulsion Laboratory) with DRIE (Deep Reactive Ion Etching) on high and low resistivity silicon wafers \cite{Jung-Kubiak17}. 
Low-resistivity wafers are not suitable for optical testing due to in-band absorption loss, but were useful for developing and verifying the fabrication process. We have now progressed to fabricating 2-layer designs on high resistivity wafers that are suitable for optical testing.
The etching was done on 1~mm thick and 100~mm diameter silicon wafers. The dimensions of the fabricated structures are very close to the design dimensions ($\pm 3 \mu m$), as shown on fig~\ref{fig:2lay_sem}.

\section{Measurement Bench}

\begin{figure}[htbp]
\begin{center}
\includegraphics[width=12cm]{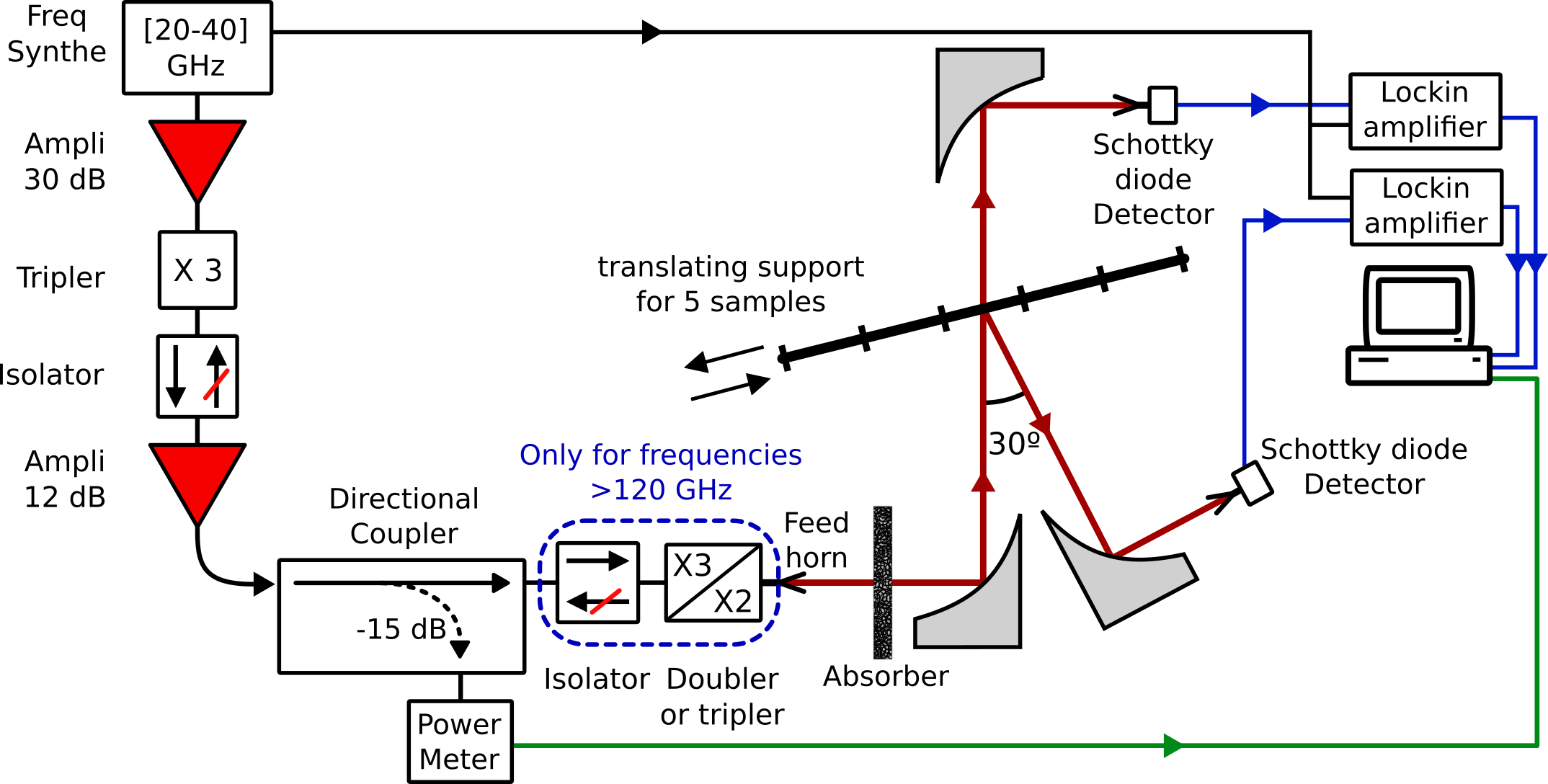}
\caption{\label{fig:schem}Schematic of the test setup to characterize the silicon wafers}
\end{center}
\end{figure}

The measurement bench (cf. fig~\ref{fig:schem}) is made of a solid-state frequency multiplier chain generating a signal (Gaussian beam) between 75~GHz and 330~GHz, a quasi-optical setup with 3 parabolic mirrors and a translating support holding the samples, and two Schottky detectors which receive the signals transmitted and reflected by the sample. 
To cover a frequency between 75 GHz and 330 GHz, we have to measure three frequency bands separately (75-115~GHz, 150-230~GHz, and 225-345~GHz). For band 1 (75-115~GHz), the frequency of the synthesizer is only multiplied by a tripler, for band 2  a doubler is added at the end of the chain, and for band 3 the doubler is replaced by a tripler. 
The signal generated by the frequency synthesizer is modulated (in amplitude) at 10~Hz which allows the use of lockin amplifiers to read the signals at the output of the Schottky detectors. The translating support makes a 15$^o$ angle with the incident beam and has five slots, three are used for samples, one is a hole and one is a mirror. To measure the transmission of the samples, we calculate the ratio of the power transmitted by the hole and by the sample, and the reflection is the ratio of the power reflected by the sample and by the mirror. A 10 dB foam absorber has been installed after the first parabolic mirror to reduce the standing waves.

\section{Single AR layer wafers: Test Results}

\begin{figure*}[htbp]
    \centering
    a) Single sided, single layer, square posts
    
    \begin{subfigure}[t]{0.45\textwidth}
        \centering
        \includegraphics[height=1.15in]{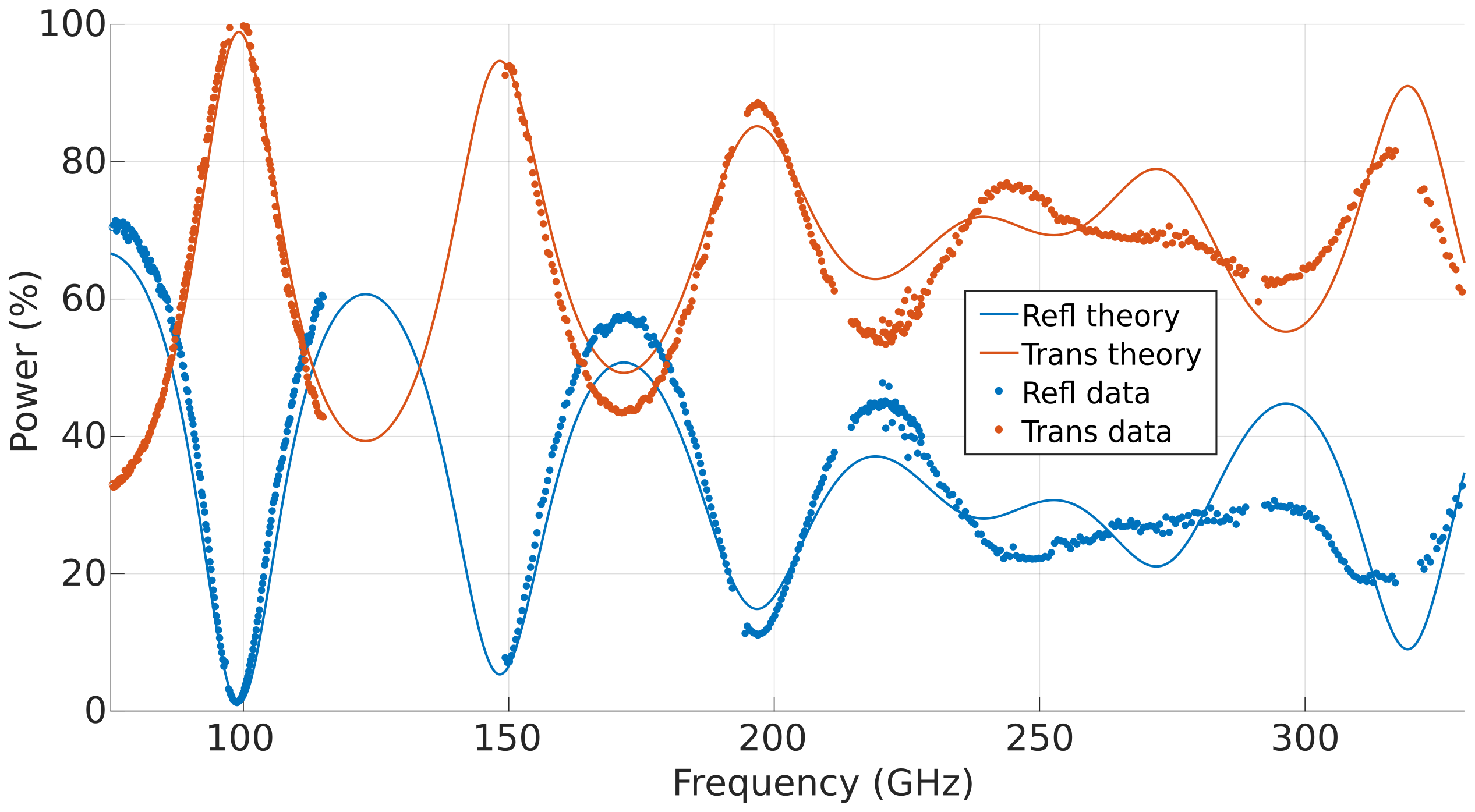}
 %       \caption{Lorem ipsum}
    \end{subfigure}%
    ~ 
    \begin{subfigure}[t]{0.45\textwidth}
        \centering
        \includegraphics[height=1.15in]{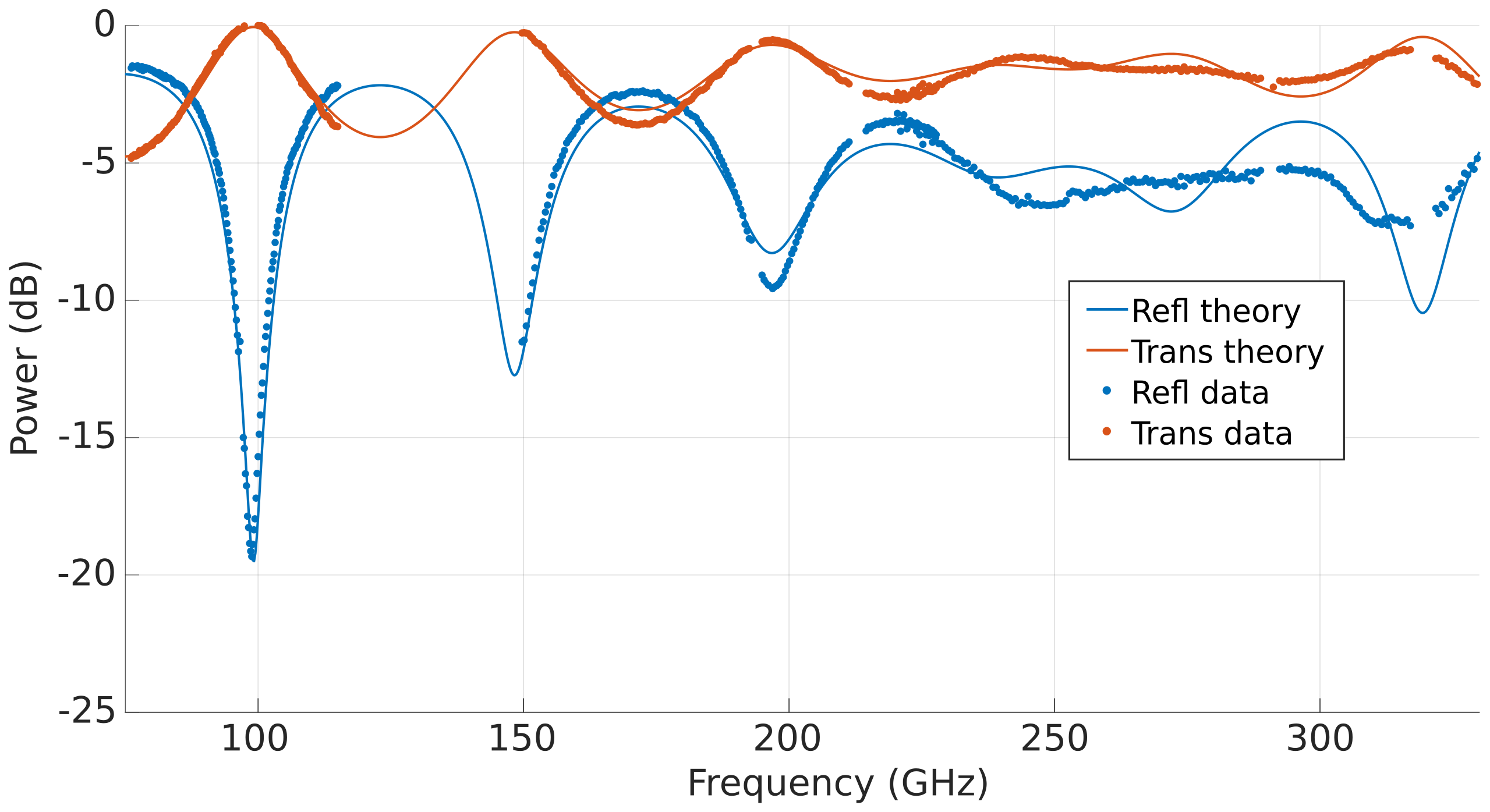}
 %       \caption{Lorem ipsum, lorem ipsum,Lorem ipsum, lorem ipsum,Lorem ipsum}
    \end{subfigure}
%     \caption{\label{fig:ss_post}Measured transmission and reflection of a single sided, single layer, posts pattern wafer, compared with HFSS simulations. {\it Left,} linear plot and {\it Right,} log plot}
\vspace{0.3cm}

b) Single sided, single layer, square holes

    \begin{subfigure}[t]{0.45\textwidth}
        \centering
        \includegraphics[height=1.15in]{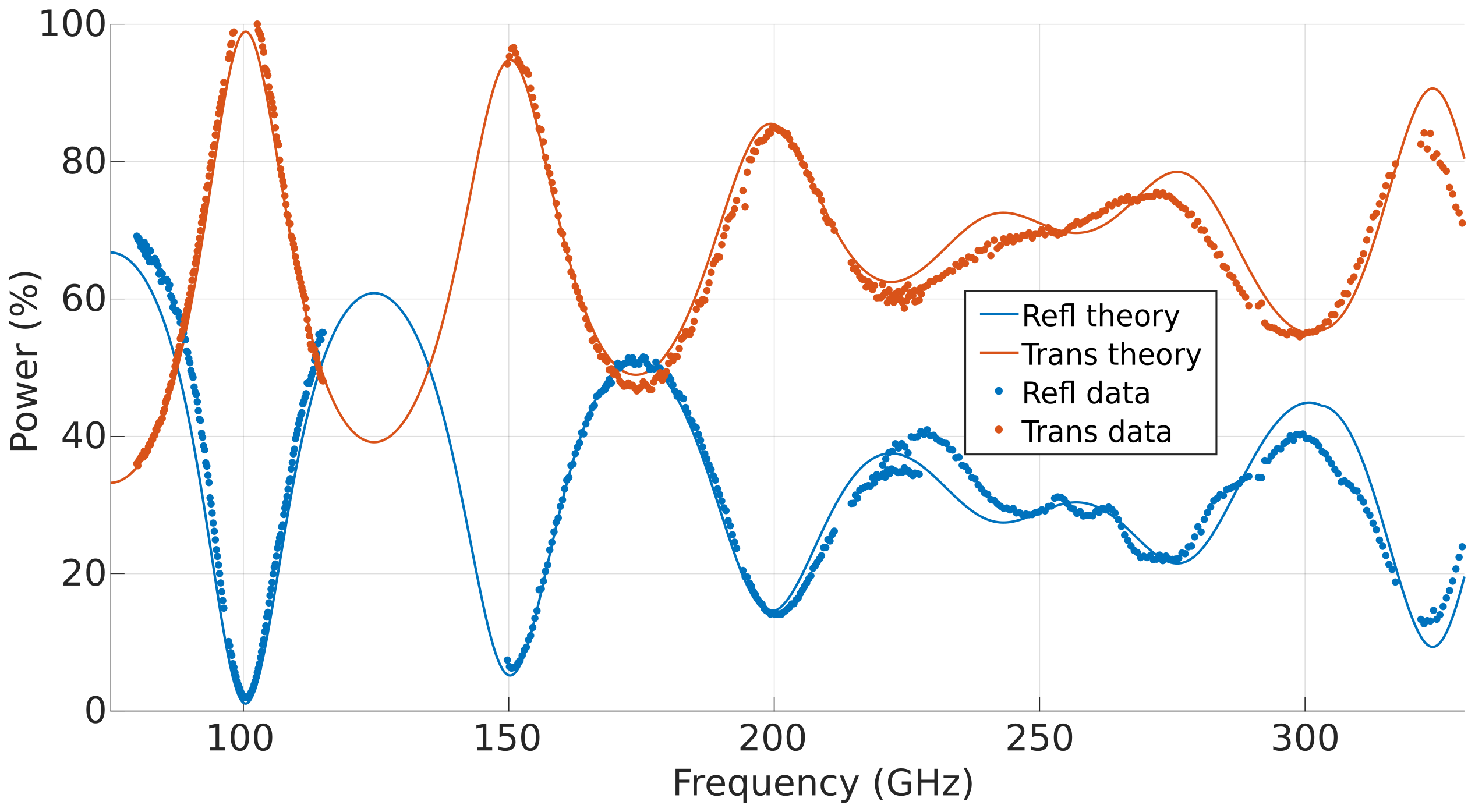}
 %       \caption{Lorem ipsum}
    \end{subfigure}%
    ~ 
    \begin{subfigure}[t]{0.45\textwidth}
        \centering
        \includegraphics[height=1.15in]{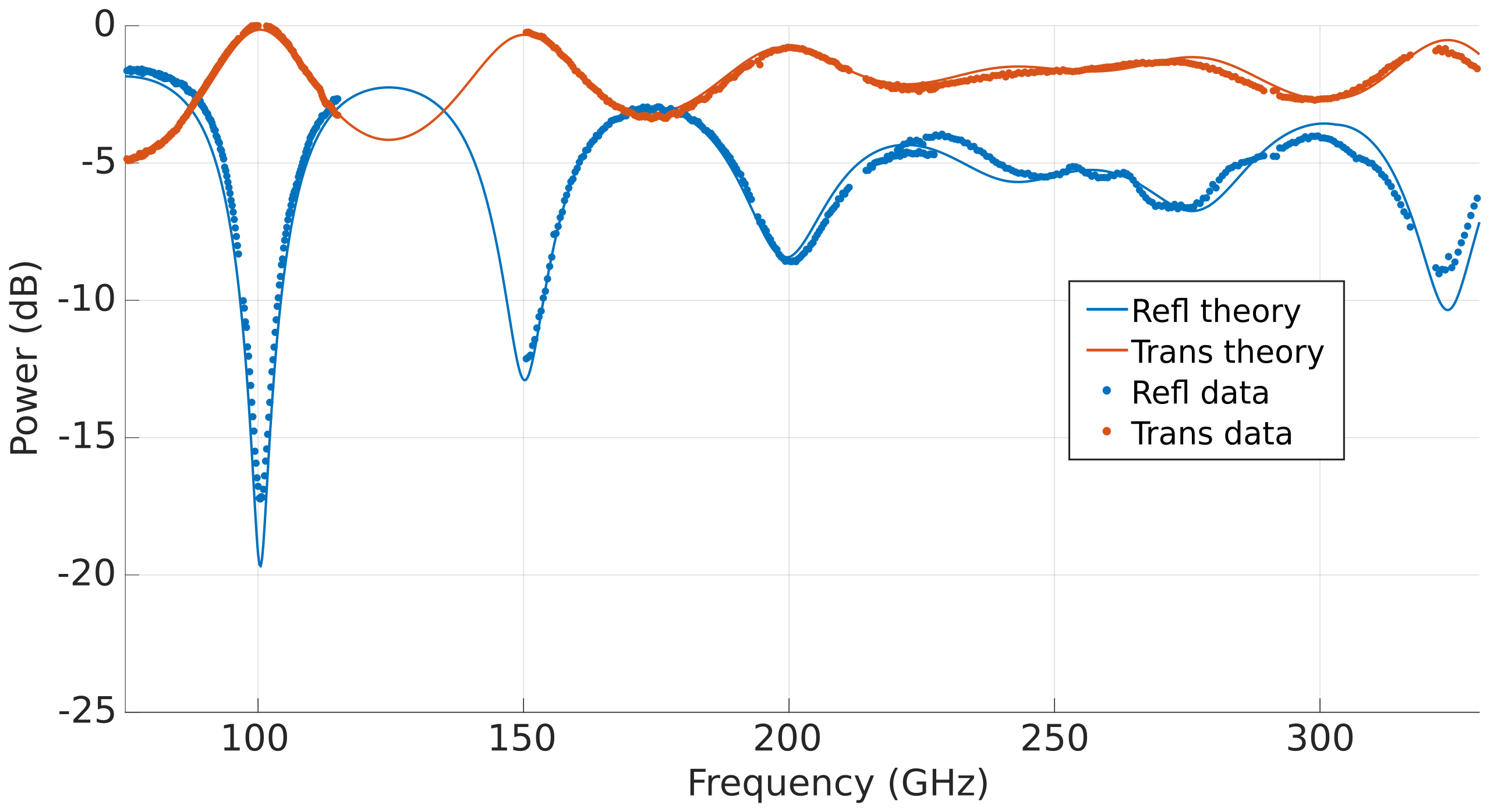}
 %       \caption{Lorem ipsum, lorem ipsum,Lorem ipsum, lorem ipsum,Lorem ipsum}
    \end{subfigure}
%     \caption{\label{fig:ss_hole}Measured transmission and reflection of a single sided, single layer, holes pattern wafer, compared with HFSS simulations. {\it Left,} linear plot and {\it Right,} log plot}
\vspace{0.3cm}

    c) Double sided, single layer, square holes
    
    \begin{subfigure}[t]{0.45\textwidth}
        \centering
        \includegraphics[height=1.15in]{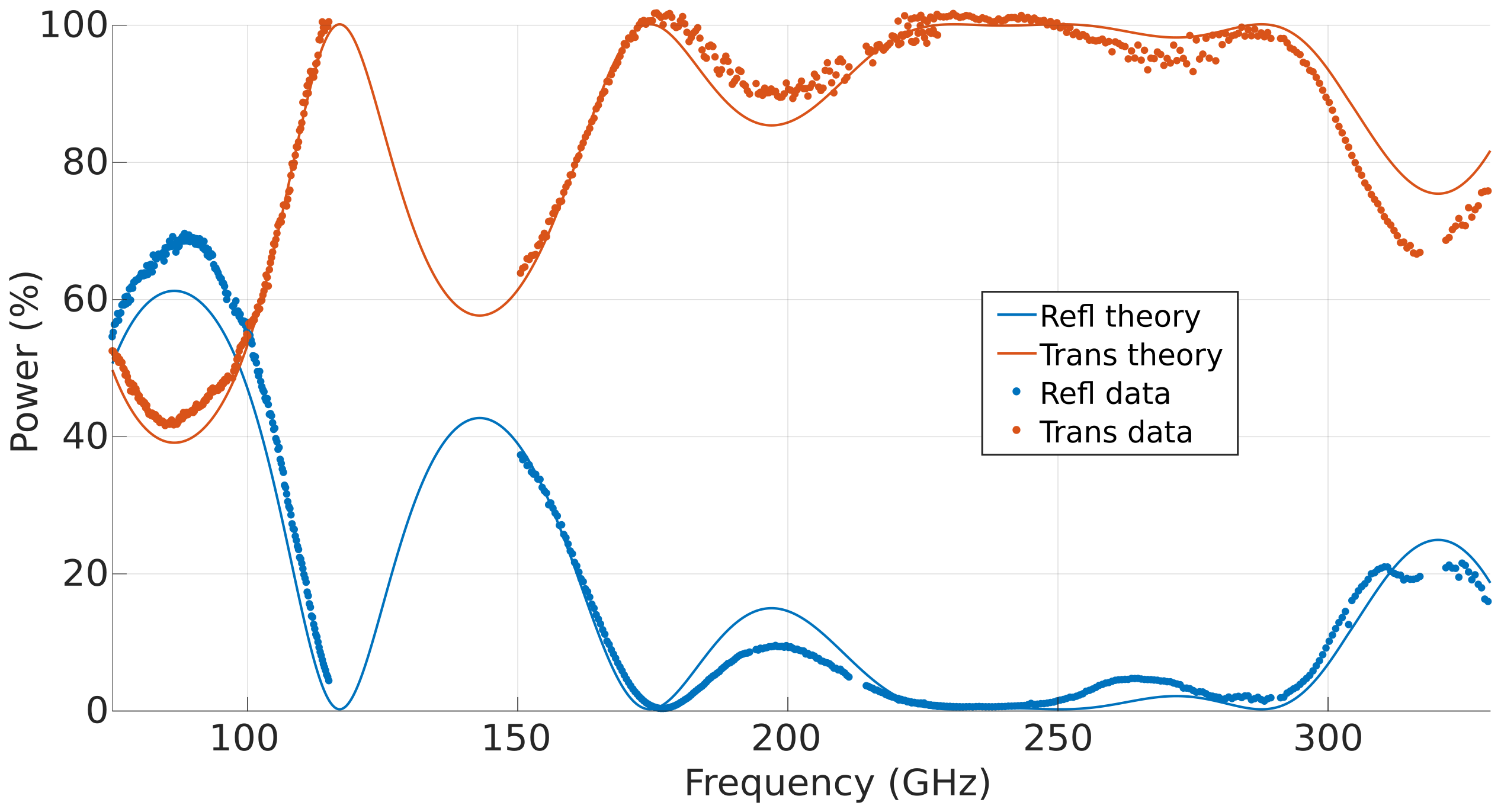}
 %       \caption{Lorem ipsum}
    \end{subfigure}%
    ~ 
    \begin{subfigure}[t]{0.45\textwidth}
        \centering
        \includegraphics[height=1.15in]{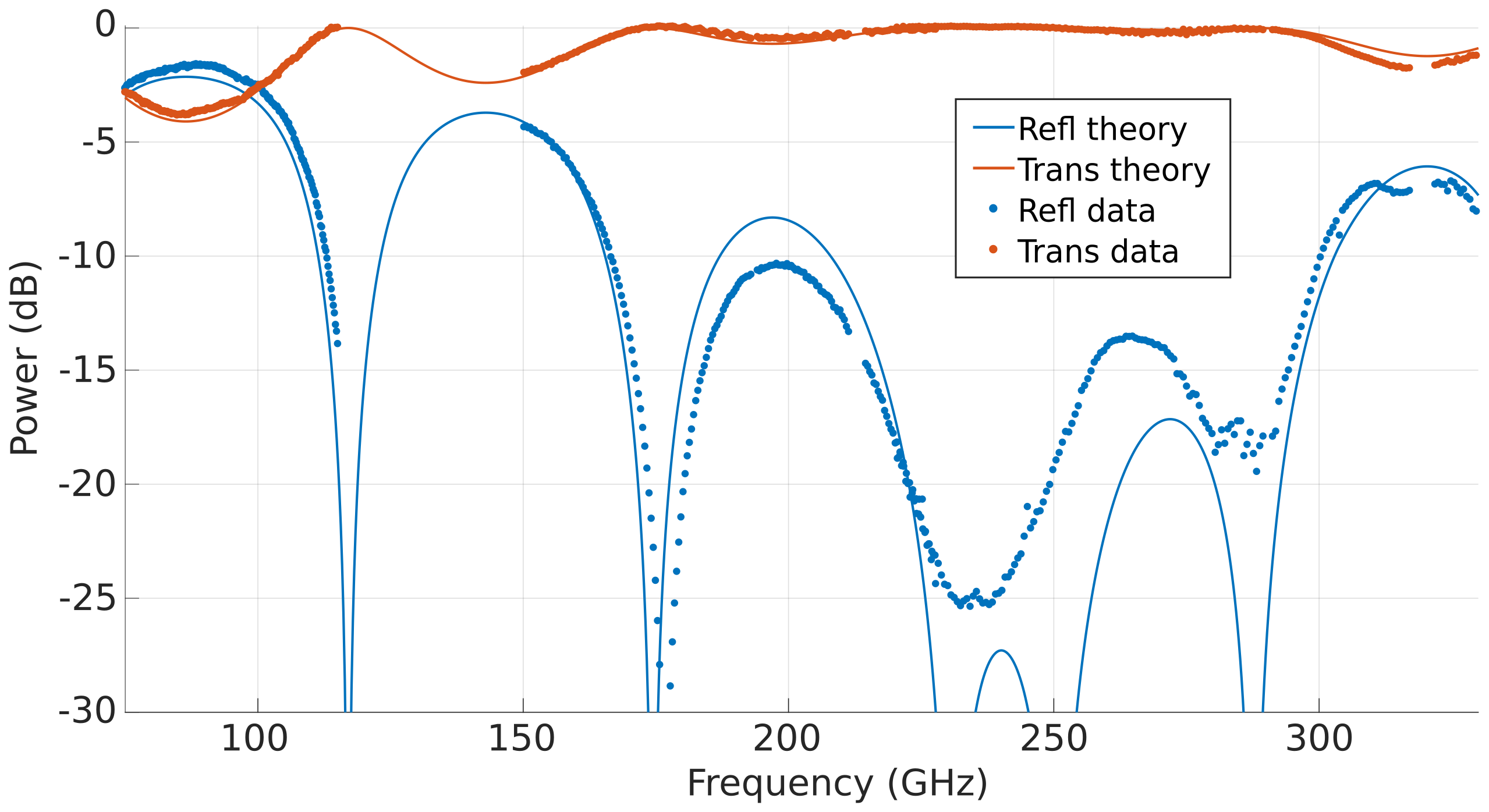}
 %       \caption{Lorem ipsum, lorem ipsum,Lorem ipsum, lorem ipsum,Lorem ipsum}
    \end{subfigure}
%     \caption{\label{fig:ds_hole}Measured transmission and reflection of a double sided, single layer, holes pattern wafer, compared with HFSS simulations. {\it Left,} linear plot and {\it Right,} log plot}
\vspace{0.3cm}

d) Two Single sided, single layer, square posts wafers bonded together

    \begin{subfigure}[t]{0.45\textwidth}
        \centering
        \includegraphics[height=1.15in]{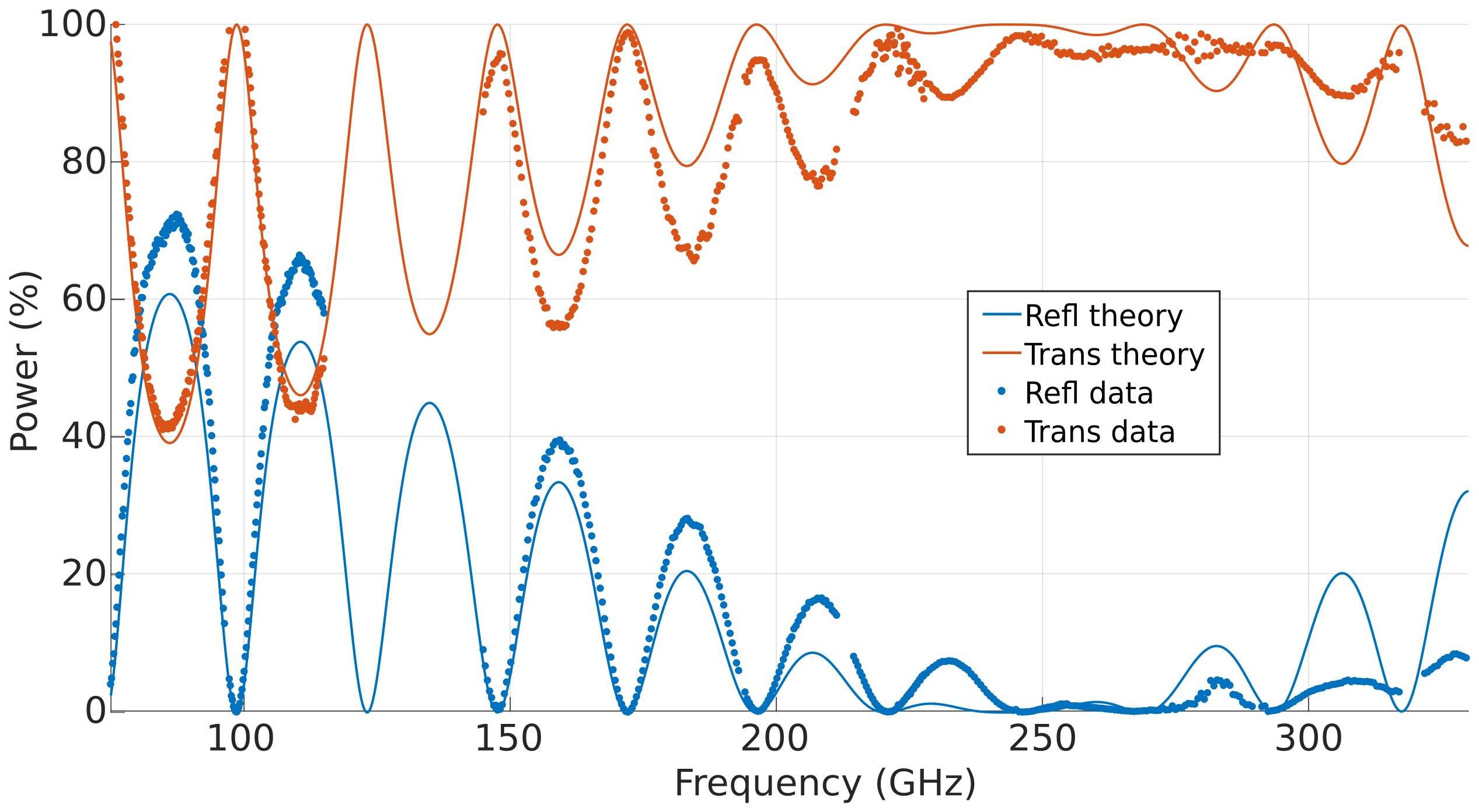}
 %       \caption{Lorem ipsum}
    \end{subfigure}%
    ~ 
    \begin{subfigure}[t]{0.45\textwidth}
        \centering
        \includegraphics[height=1.15in]{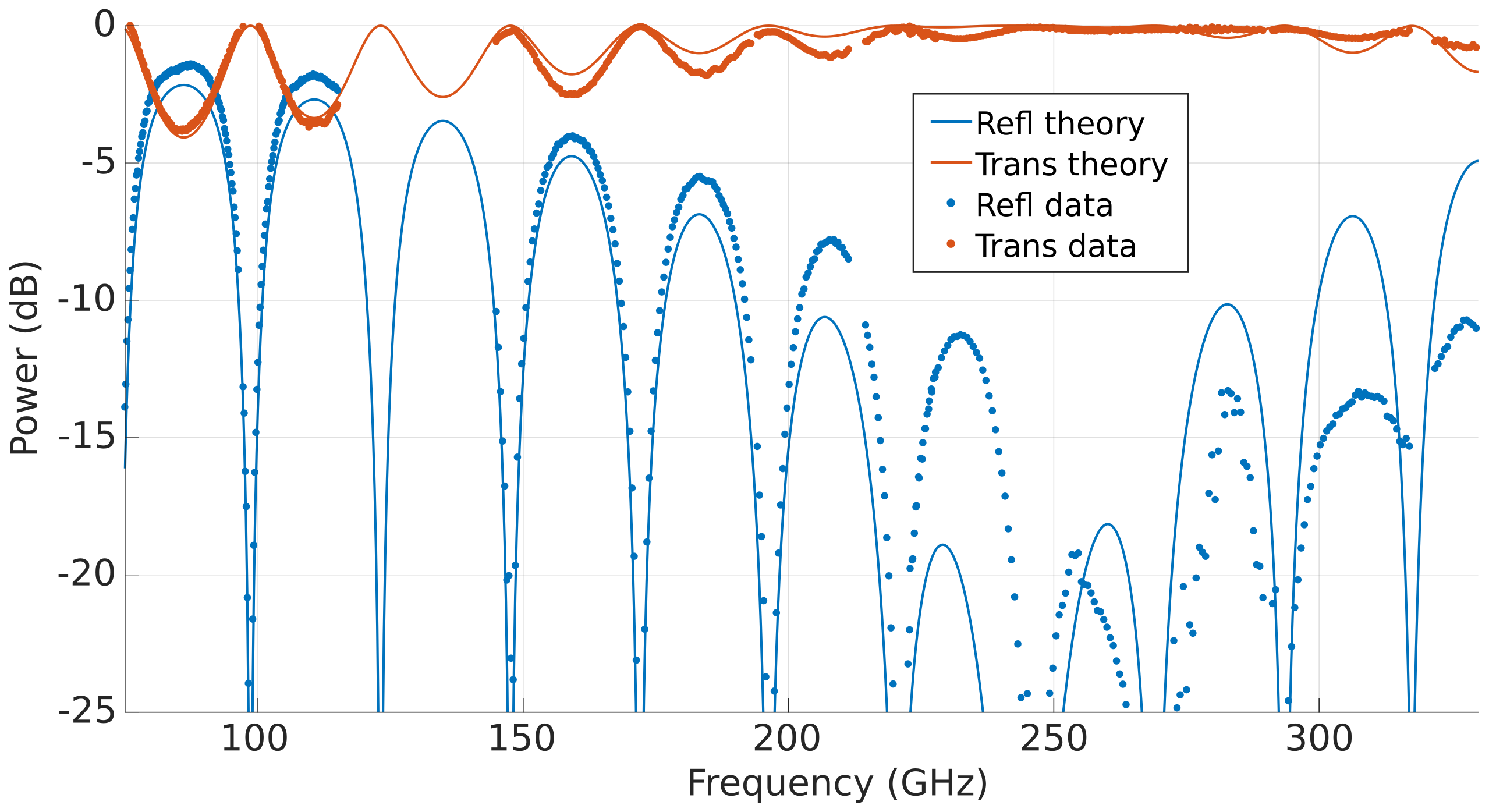}
 %       \caption{Lorem ipsum, lorem ipsum,Lorem ipsum, lorem ipsum,Lorem ipsum}
    \end{subfigure}
%     \caption{\label{fig:2xss_post}Measured transmission and reflection of a two bonded single sided, single layer,  posts pattern wafer, compared with HFSS simulations. {\it Left,} linear plot and {\it Right,} log plot}
\vspace{0.3cm}

e) Two Single sided, single layer, square holes wafers bonded together

    \begin{subfigure}[t]{0.45\textwidth}
        \centering
        \includegraphics[height=1.15in]{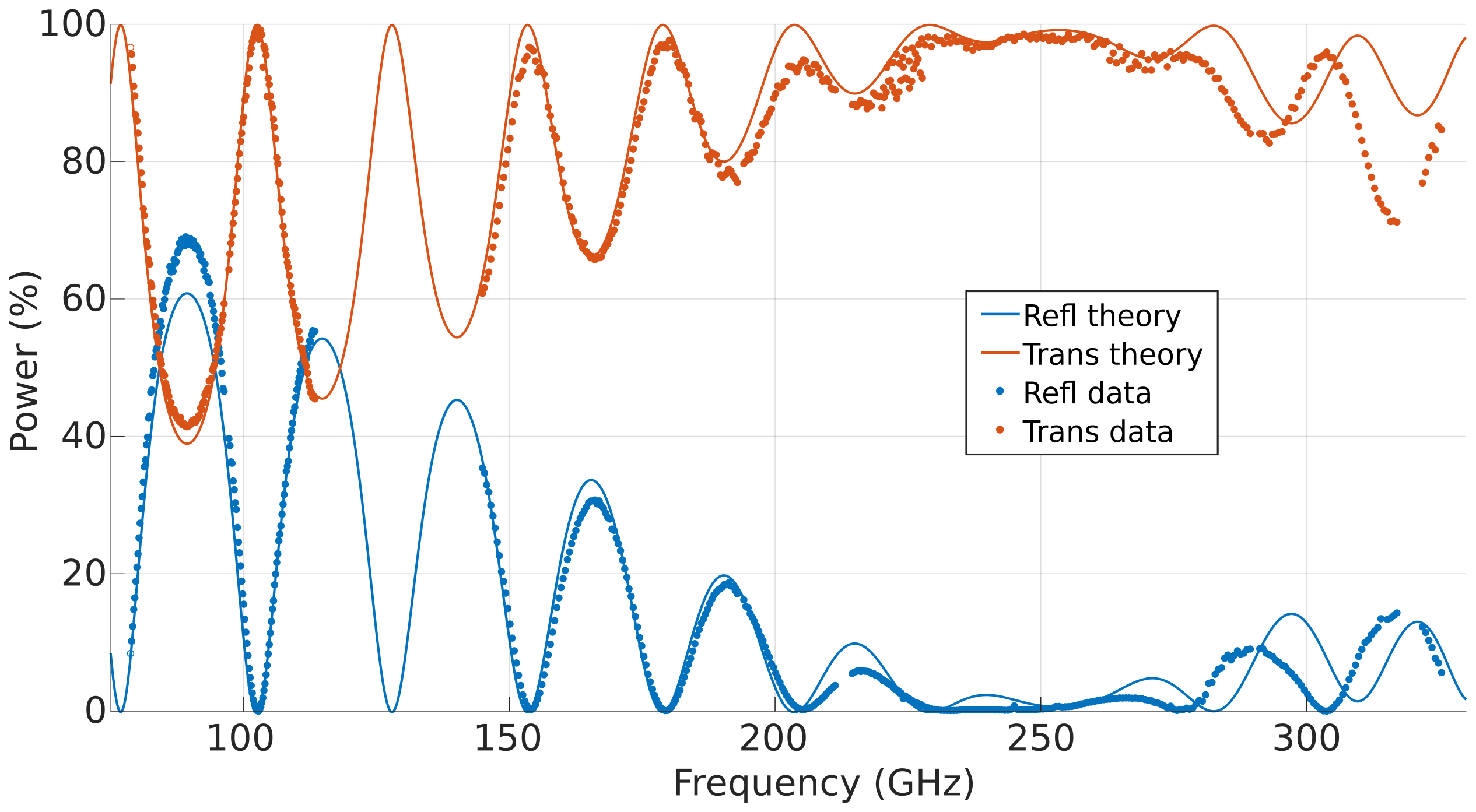}
 %       \caption{Lorem ipsum}
    \end{subfigure}%
    ~ 
    \begin{subfigure}[t]{0.45\textwidth}
        \centering
        \includegraphics[height=1.15in]{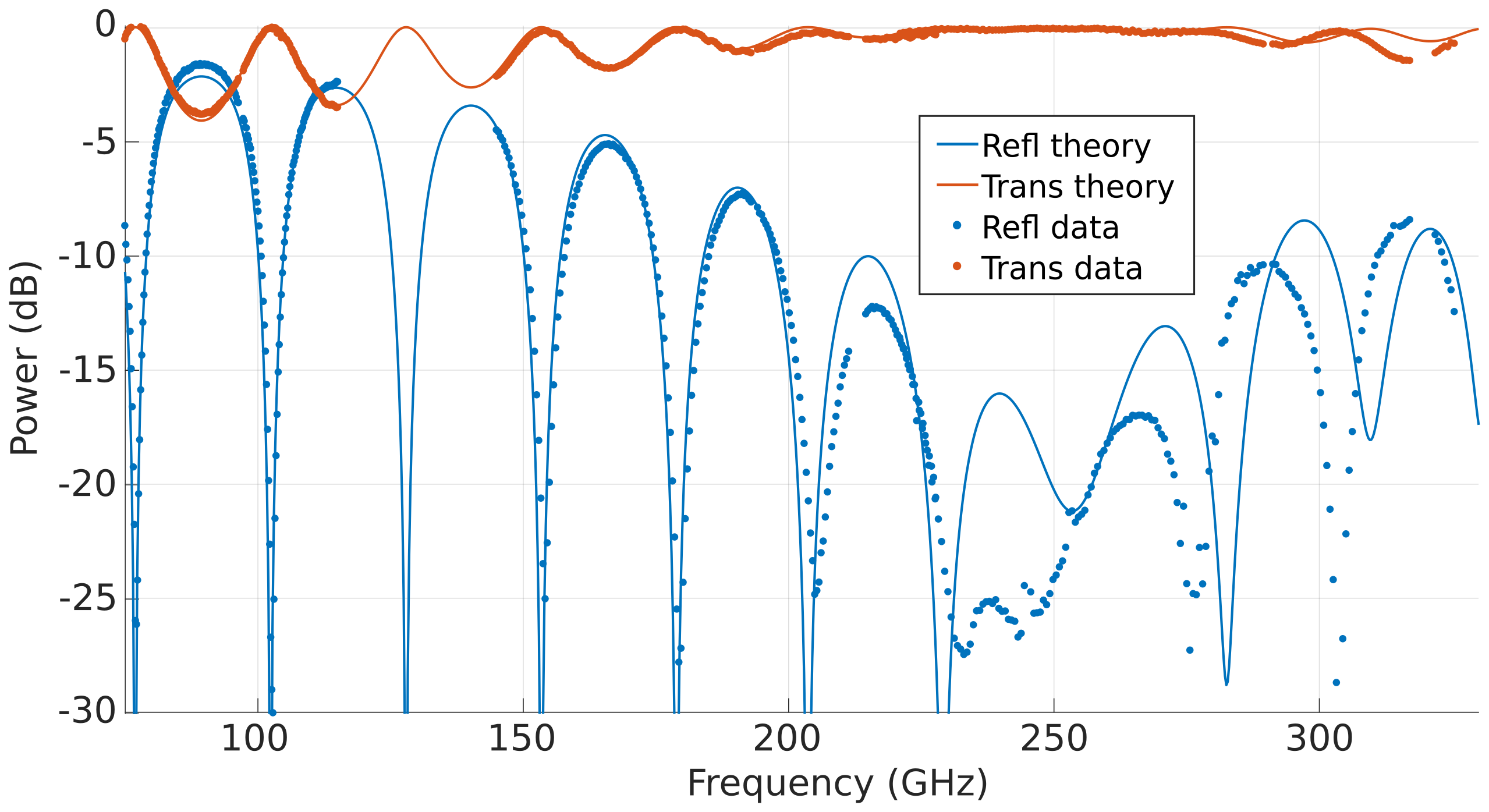}
 %       \caption{Lorem ipsum, lorem ipsum,Lorem ipsum, lorem ipsum,Lorem ipsum}
    \end{subfigure}
%      \caption{\label{fig:2xss_hole}Measured transmission and reflection of a two bonded single sided, single layer,  holes pattern wafer, compared with HFSS simulations. {\it Left,} linear plot and {\it Right,} log plot}
     \caption{\label{fig:Meas}Measured transmission and reflection of high resistivity silicon single layer wafers compared with HFSS simulations, {\it Left,} linear plots and {\it Right,} log plots}. 
\end{figure*}

Only the single-layer high resistivity silicon wafers have been tested (the double-layer samples will be tested soon). Three different configurations were fabricated and tested: single sided wafers with square holes and posts (cf.~figs.~\ref{fig:Meas}~a~\&~b), double sided wafers with square holes (cf.~fig.~\ref{fig:Meas}~c), two single sided wafers bonded together with square holes and posts (cf.~figs.~\ref{fig:Meas}~d~\&~e). For each of these 3 configurations and for the 2 geometries (holes and posts), the test results have been compared with the simulation results given by HFSS. 

\section{Conclusion}

The tests of single-layer AR coatings agree well with the HFSS simulations, which validates the fabrication and testing setup. 
The transmission of the tested single-layer double sided sample is better than 90\% between 170~GHz and 330~GHz. However, to study the polarization of the CMB and the galaxy clusters, we need a transmission better than 99\% between 75~GHz and 420~GHz, which will require a 7 layer AR coating. Double-layer samples have already been successfully fabricated, and new designs with more layers, incorporating wafer-bonding, are planned to be fabricated in a near future. Finally, our goal is to combine this broadband multilayer AR coating with gradient index focusing optics.

\begin{acknowledgements}
This work was funded by NASA (Grant NNX15AE01G).
\end{acknowledgements}

\pagebreak

\end{document}